\documentclass{optica-article}

\journal{opticajournal} 

\articletype{Research Article}

\usepackage{lineno}
\usepackage{textcomp}
\usepackage{listings}
\begin{document}

\title{Versatile, open-source program for simulating high-harmonic generation}

\author{Christian A. Schröder\authormark{1,2,3,*}, Reinhard Kienberger\authormark{3} and Stephen R. Leone\authormark{1,2}}

\address{\authormark{1}Department of Chemistry, University of California, Berkeley, CA 94720, USA.\\
\authormark{2} Chemical Sciences Division, Lawrence Berkeley National Laboratory, Berkeley, CA 94720, USA.\\\authormark{3}Chair for Laser- and X-ray physics E11, Department of Physics, TUM School of Natural Sciences, Technische Universität München, James-Franck-Str. 1,85748 Garching}

\email{\authormark{*}caschroeder@lbl.gov} 


\begin{abstract*}
	Light sources based on high-harmonic generation (HHG) underpin ultrafast spectroscopy experiments across a large range of photon energies, spanning from the extreme-ultraviolet to the soft x-ray. To this day their design, implementation and improvement presents uniqe challenges, but can be aided by numerical tools. Here we present a new simulation program designed for this purpose, which accurately takes both macroscopic and microscopic aspects of high-harmonic generation into account and is therefore applicable across the broad range of parameters that HHG based light sources are today utilized in. The program is validated by calculating harmonic emission in four common experimental configurations.
\end{abstract*}

\section{Introduction}

High-harmonic generation (HHG), i.e. the upconversion of highly intense low-frequency laser light via strongly non-pertutbative interaction with a gaseous medium \cite{ferray1988multiple, li1989multiple}, has driven groundbreaking developements in ultrafast spectroscopy, and to this day remains the ultimate table-top method for generating femtosecond and attosecond bursts of extreme ultraviolet (XUV) and soft x-ray (SXR) radiation. Due to its utility and versatility, both the macroscopic and microscopic aspects of efficiently generating said pulses have received a great deal of theoretical and experimental attention (see e.g. \cite{lewenstein1994theory, balcou1997generalized, gaarde2008macroscopic, heyl2016scale, johnson2018high} and references therein). While analytical approaches have yielded fundamental insights into how the harmonic emission can be controlled spatially, spectrally and temporally, direct numerical simulation of the macroscopic HHG remains very important due to the intricate and complex interaction of linear- and non-linear effects in real-world HHG sources that may render an analytical approach infeasible, especially when intensities and gas pressures are high. Here we present a program for simulating HHG that will aid experimentalists in designing and deploying HHG sources across a large range of driving laser parameters and gas pressures. The approach sets itself apart from other concurrent simulation tools by being monolithic (meaning that no interfacing between different programs or modules is needed, in contrast to e.g. \cite{vabek_thesis, vabek_proceedings}), minimizing the amount of post-processing, and furthermore striving for applicability in both the high- and low gas-pressure regimes (as opposed to e.g. \cite{hernandezgarcia2010high, hhgmax}).

\section{Computational model}

We follow \cite{gaarde2008macroscopic} and solve the propagation equation (\ref{eq:propagation}) in a reference frame traveling with a laser pulse in free space,
\begin{align}
	\nabla^2_\perp E_x(\omega) +\frac{2 i\omega}{v_\mathrm{g}(\omega)}\frac{\partial E_x(\omega)}{\partial z}
	+ \left[\frac{n^2(\omega)\omega^2}{c^2} - \frac{\omega^2}{v_\mathrm{g}(\omega)^2}\right] E_x(\omega) = -\frac{\omega^2}{\varepsilon_0 c^2}P_{\mathrm{nl}, x}(\omega),
	\label{eq:propagation}
\end{align}
where for  $x=f$ above equation describes the propagation of the fundamental field, and for $x = h$ it describes the harmonic field. The radial coordinate $r$ and axial coordinate $z$ are dropped in our notation for brevity, but all electric fields $E_x(\omega)$ and polarizations $P_{\mathrm{nl}, x}(\omega)$ are understood to depend on $r$ and $z$. The fundamental driving field $E_f(\omega)$ and the harmonic field $E_h(\omega)$ are propagated separately but in lockstep and are coupled in each $z$-plane via the macroscopic non-linear polarization $P_{\mathrm{nl}, h}(\omega)$ induced by the fundamental calculated using the Lewenstein model \cite{lewenstein1994theory}. For the fundamental, on the other hand, $P_{\mathrm{nl}, f}(\omega)$ accounts for the medium's third-order nonlinearity as well as a contribution due to plasma generated by the high-intensity laser field, but no back-action of the harmonic radiation on the driving field is allowed as frequency mixing processes of the driving pulse with the weak harmonic field are expected to be negligible. The linear optical properties (refraction and extinction) of the gas medium enter eq. \ref{eq:propagation} via the complex refractive index $n(\omega)$ and the group velocity $v_g(\omega)$, which may be $z$-dependent to allow for different gas-cell designs. Sellmeier-type formulae are used to calculate $n(\omega)$ for the fundamental \cite{Boerzsoenyi2008}, while tabulated values (e.g. from \cite{henke1991xray}) are interpolated for the harmonic field. Neumann boundary conditions are enforced with first-order accuracy at $r = r_\mathrm{max}$ and second-order accuracy at $r = 0$. An electric field $\left.E_f(\omega)\right|_{z = z_0} = E_f^{(0)}(\omega)$ is specified as an initial condition for the fundamental, while $\left.E_f(\omega)\right|_{z = z_0} = 0$ is assumed for the harmonics. The main results of a calculation are the final complex electric fields $E_h(\omega, r)$ and $E_f(\omega, r)$ of the harmonic and fundamental field respectively at the end of the simulation region. From these the final harmonic spectrum can be evaluated e.g. on-axis as $\left|E_h(\omega, r)\right|^2$, or alternatively by integrating over the radial coordinate,
\begin{equation}
	S(\omega) \propto \int_0^\infty\left|E_h(\omega, r)\right|^2rdr.
\end{equation}
In the examples below, radially integrated spectra are reported throughout.

\subsection{Source terms}

\subsubsection{Fundamental field}

Self-focusing and spectral broadening of the fundamental beam due to the Kerr-effect enters the non-linear polarization on the righ hand side of eq. \ref{eq:propagation} as
\begin{equation}
	P_\mathrm{Kerr}(t) = \rho(z)\bar\gamma^{(3)}\left|E_f(t)\right|^2E_f(t),
	\label{eq:kerr}
\end{equation}
where $\bar\gamma^{(3)}$ is the third-order hyperpolarizability available from literature (e.g. \cite{lehmeier1985nonresonant}), and $\rho(z)$ is the gas density. Plasma formed in the interaction region due to ionization will contribute to the non-linear polarization of the medium via currents driven by the fundamental field. The associated current density $j(t)$, which relates to the non-linear polarization as $-\omega^2 P_\mathrm{p}(\omega) = \mathcal{F}\left\{\partial j(t)/\partial t\right\}$ \cite{gaarde2008macroscopic} is taken to include two contributions,
\begin{align}
	j_\mathrm{p}(t)&=n_e(t)E_f(t),\\
	j_\mathrm{abs}(t)&=\frac{\partial}{\partial t}\left(\frac{1}{\left|E_f(t)\right|}\sum_i n_i(t)\gamma_i(t)I_p^{(i)}\right),
\end{align}
where $n_e(t)$ is the free electron density, $n_i(t)$ is the density of $i$-fold ionized gas atoms, $I_\mathrm{p}^{(i)}$ is the $i$-th ionization potential and $\gamma_i(t)$ is the ionization rate for the $i$-th ionic state. The spatial dependence is again dropped for brevity in all terms. The plasma oscillation term $j_\mathrm{p}(t)$ is responsible for defocusing and blue-shifting of the driving field, while $j_\mathrm{abs}(t)$ ensures that Ohmic energy dissipation is balanced with energy loss due to ionization. The rates $\gamma_i(t)$ are determined from the model of Ammosov, Delone and Krainov \cite{adk} from the instantaneous values of the electric field $E_f(t)$. Nonadiabatic corrections \cite{gaarde2006spatial} can be switched on if desired. The $n_i(t)$ are determined from a set of coupled rate equations
\begin{equation}
	\frac{d}{dt}n_i(t) = - \gamma_i(t) n_i(t) + \gamma_{i - 1}(t) n_{i - 1}(t),
	\label{eq:ionisation_rate_eq}
\end{equation}
and the electron density is taken as $n_e(t) = \sum_i i n_i(t)$.

\subsubsection{Harmonic field}
The atomic dipole response $x(t)$ for high-harmonic emission is calculated via Lewenstein's model \cite{lewenstein1994theory, antoine1996theory},
\begin{align}
	x(t) = 2\Re\left\{
		i\int_0^\infty
		d^{*}\left(p_\mathrm{st}(t, \tau) - A_f(t)\right)
		d\left(p_\mathrm{st}(t, \tau) - A_f(t - \tau)\right)
		E_f(t - \tau)
		e^{-iS(t, \tau)}
		 d\tau\right\},
 		\label{eq:lewenstein}
\end{align}
with the stationary momentum $p_\mathrm{st}(t, \tau) = \int_{t - \tau}^{t}dt''A_f(t'')/\tau$ and the action
\begin{equation}
	S(t, \tau) = \int_{t - \tau}^{t}dt''\left(\frac{1}{2}\left(p_\mathrm{st}(t, \tau) - A_f(t'')\right) + I_p\right).
	\label{eq:action}
\end{equation}
All throughout we use the hydrogenic transition dipole moment
\begin{equation}
	d(p) = i\frac{2^\frac{7}{2}(2I_p)^\frac{5}{4}}{\pi}\frac{p}{(p^2 + 2I_p)^3}.
	\label{eq:hydrogenic_dp}
\end{equation}
In eqns. \ref{eq:lewenstein}, \ref{eq:action} and \ref{eq:hydrogenic_dp} $I_p$ is the ionization potential of the medium and $A_f(t)$ is the vector potential corresponding to the fundamental's electric field $E_f(t)$. Following \cite{gaarde2008macroscopic} the macroscopic polarization is then
\begin{equation}
	P_\mathrm{nl, h}(t) = x(t)n_0(t)\rho(z),
\end{equation}
with $n_0(t)$ being the fraction of neutral atoms at time $t$.

\section{Implementation}
The propagation equation (eq. \ref{eq:propagation}) is solved using a Crank-Nicholson integrator. It is linearized by computing the nonlinear source terms in the time domain from the current fields $E_x(t)$. Initial values for the fundamental field $E_f(\omega)$ are given in the frequency domain, either as a data file tabulating a power spectrum and spectral phase, or a central frequency, spectral width and amplitude in which case a Gaussian spectrum with these parameters is set up, for which polynomial coefficients for the spectral phase can be supplied. Multiple Gaussian spectra can be defined at the same time, permitting e.g. the study of HHG driven by multi-color fields, with their spectral phases being defined independently. The time and frequency grid in the calculation are related by the standard Fourier relations, where extent and resolution are set by the user: Nyquist frequency (maximum photon energy in the calculation, given in electron volt) and number of points of the frequency axis are determined by the user. The maximum energy should be slightly higher than the highest harmonic signal expected in the calculation. Currently only Gaussian beam profiles characterized solely by their beam waist $w_0$ in vacuum are implemented for the fundamental as these present the most common case, but Laguerre-Gaussian beams will be implemented in the future. The cylindrical spatial grid can freely be adjusted in extent and resolution. The optical properties of the gas medium are given with respect to a reference pressure and temperature and scaled appropriately. The gas-density profile $\rho(z)$ can be set up by the user to simulate various gas-cell designs and is given as a list of position--pressure ratio point pairs $(z_i, f_i)$. The pressure at a point $z$ is then $p(z) = f p_0$ where $p_0$ is the maximum pressure, $f$ is calculated via a linear interpolation from the given point pairs, and the density is calculated from pressure and temperature assuming an ideal gas.

If appropriate, the integration of eq. \ref{eq:lewenstein} can be restricted to smaller time windows (e.g. only over one period of the driving field), resulting in a massive speed-up of the calculation. Additionally, in cases where the fundamental beam profile varies slowly with $z$ it may not be necessary to calculate $x(t)$ at every step, instead the program can be set up to only do so in every $n$-th plane, interpolating the source term for intermediate $z$-positions, drastically cutting down calculation time. Eqns. \ref{eq:ionisation_rate_eq} require integration with a stepsize-adaptive method for which we choose the explicit fourth-order Runge-Kutta scheme.

The simulation program is implemented in \texttt{C++} and uses the \texttt{armadillo} library \cite{sanderson2019practical, sanderson2025armadillo} for arrays and general math and \texttt{fftw3} \cite{FFTW05} for fast-Fourier-transforms. The source code is available openly in a \texttt{git} repository (\url{https://gitlab.com/caschroeder/hhg}) together with comprehensive documentation. A calculation is configured using a simple text file, lending itself to automation of parameter scans e.g. via \texttt{UNIX}-shell-scripting. A workstation computer running some flavor of Linux is recommended. All of the examples shown were run on a workstation computer with an Intel\textsuperscript{\tiny\textregistered} Xeon\textsuperscript{\tiny\textregistered} W-2265 CPU clocked at $3.5\,\mathrm{GHz}$ or a Lenovo\textsuperscript{\tiny\textregistered}  ThinkPad\textsuperscript{\tiny\textregistered} T430 with an  Intel\textsuperscript{\tiny\textregistered} i5 CPU running at $2.6\,\mathrm{GHz}$. Both computers run recent versions of the Debian GNU/Linux operating system. The integrator is currently not parallelized. Upon completion of a calculation, the final complex electric fields can be written into a \texttt{hdf5} file. A plain-text report on the calculation including the final radially resolved harmonic spectrum is also produced on the standard output, and additional information (e.g. the evolution of the driving field or intensity distribution) can be written to plain text files if so desired.

\section{Examples}
Here we show some example calculations to demonstrate the validity and versatility of our simulation program.

\subsection{Analysis of phase-matching conditions}
\label{ex:1}
\begin{figure}[t!]
	\includegraphics[width=\textwidth]{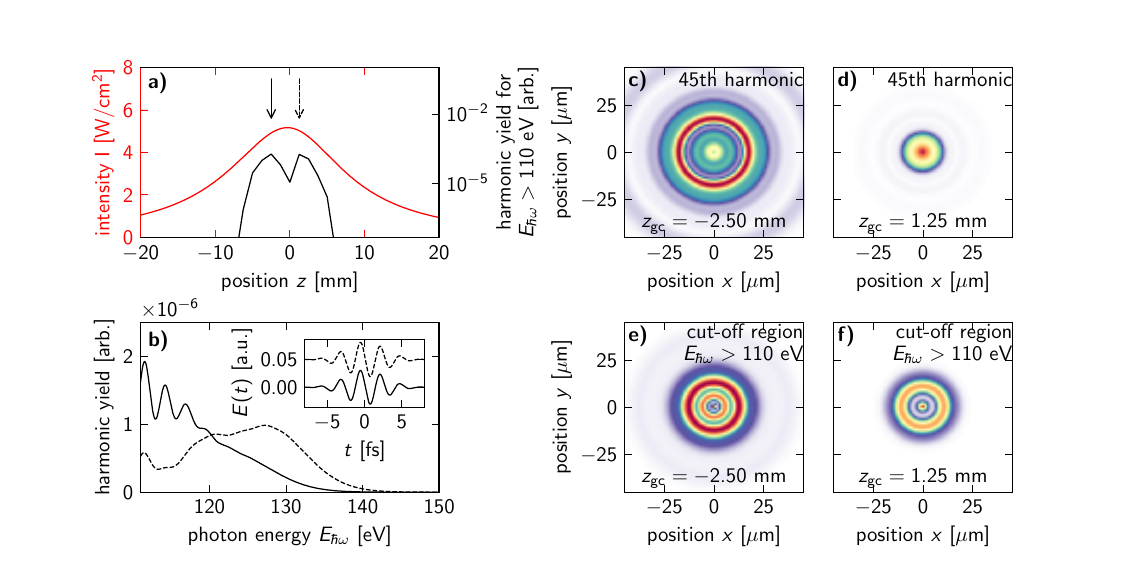}
	\caption{Analysis of phase-matching conditions with a few-cycle pulse at a driving wavelength of $800\,\mathrm{nm}$. When a short gas cell ($0.5\,\mathrm{mm}$, backed with $100\,\mathrm{mbar}$ of Helium) is moved across the focus of the driving beam the conversion efficiency for the cut-off harmonics ($E_{\hbar\omega} > 110\,\mathrm{eV}$) goes through two distinct maxima on opposite sides of the focal point, as seen in panel (a). We plot the harmonic spectra obtained at these two gas cell positions in panel (b), where the solid line is the spectrum obtained for the gas cell being placed \textsl{before} the focus and the dashed line is the spectrum for the gas cell placed \textsl{after} the focus (c.f. the solid and dashed arrows in panel (a)). The inset shows the driving pulses, slightly offset along the ordinate for clarity. Panels (c-f) compare the resulting near-field beam profiles at these gas cell positions for the $45\mathrm{th}$ harmonic (i.e. in the plateau region) and the cut-off region.}
	\label{img:z_scan}
\end{figure}

Typically, phase-matching in high-harmonic generation is controlled by careful selection of the focusing conditions, adjustment of the intensity and spot-size with a variable aperture, as well as the gas pressure, gas cell length and gas cell position with respect to the focus position of the driving beam. Here we sweep the position of a short gas cell ($0.5\,\mathrm{mm}$ in length) backed by $100\,\mathrm{mbar}$ of Helium along the propagation axis of an $800\,\mathrm{nm}$ beam with a $40\,\mathrm{\mu m}$ radius spot size at its focus. The driving pulse is short ($\sim 7\,\mathrm{fs}$). Figure \ref{img:z_scan} summarizes this calculation and panel (a) shows how the harmonic yield in the cut-off region ($E_{\hbar\omega} > 110\,\mathrm{eV}$) goes through two distinct maxima as a function of gas cell position. The corresponding spectra and driving pulses are shown in panel (b). Additionally to the spectra, the spatial profile of the emitted harmonic radiation also changes as a function of gas-cell position as it is shown in panels (c-f) of figure \ref{img:z_scan}. While placing the gas cell before the focus results in overall greater conversion efficiency, it reduces the quality of the near-field spatial profile. Placing it after results in a smoother spectrum as well a better beam shape. The results compare well to those originally presented by Balcou \textsl{et al.} \cite{balcou1997generalized} for similar parameters.

\subsection{Attosecond pulse synthesis}
\label{ex:2}
\label{sec:attosecond_pulses}
\begin{figure}[h!]
	\includegraphics[width=\textwidth]{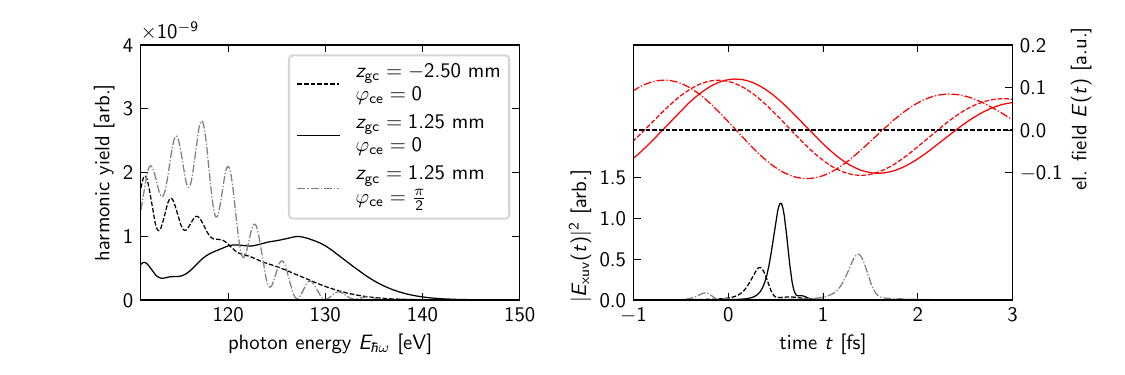}
	\caption{Simulation of isolated attosecond pulse production. In the left panel shows the on-axis harmonic emission in the spectral domain for three configurations of gas-cell position $z_\mathrm{gc}$ and carrier-envelope phase $\varphi_\mathrm{ce}$ of the driving pulse. The smoothest spectrum is generated when the gas-cell is placed slightly after the focus and $\varphi_\mathrm{ce} = 0$ (black solid line). Using a filter that rejects all photon energies below $110\,\mathrm{eV}$ a bright isolated XUV pulse of less than one femtosecond duration is synthesized in the time-domain in this case as seen in the right panel. Results for the other configurations and the respective driving fields at the gas-cell center are also shown.}
	\label{img:attosecond}
\end{figure}
Among the most spectacular applications of high-harmonic generation is the synthesis of isolated XUV/soft x-ray pulses with durations of a few hundred attoseconds first demonstrated in the early 2000s \cite{hentschel2001attosecond}. Here we show that the code is fit for the analysis conditions for the synthesis of such pulses at the example of the so-called amplitude gating scheme (cf. \cite{chini2014generation}), where a waveform-controlled few-cycle pulse generates a harmonic spectrum with a smooth cut-off region from which, via spectral filtering, a sub-femtosecond XUV pulse can be isolated.

The driving field is a $\sim 7\,\mathrm{fs}$ gaussian pulse spectrally centered at $790\,\mathrm{nm}$ focused to a spot size of $58\,\mathrm{\mu m}$ FWHM with a nominal peak intensity of $5\cdot 10^{14}\,\mathrm{W/cm^2}$. Under these conditions, using Helium as the nonlinear medium the cut-off is expected around $120\,\mathrm{eV}$ photon energy. We calculate the on-axis harmonic spectrum for three configurations of gas-cell position and carrier-envelope phase $\varphi_\mathrm{ce}$ of the driving field. The results are summarized in fig. \ref{img:attosecond}. A strong and isolated attosecond pulse can be synthesized from the cut-off portion of the harmonic spectrum when the gas-cell is placed slightly after the focus and $\varphi_\mathrm{ce} = 0$.

\subsection{Two-color driving fields}
\label{ex:3}
\begin{figure}[h!]
	\includegraphics[width=\textwidth]{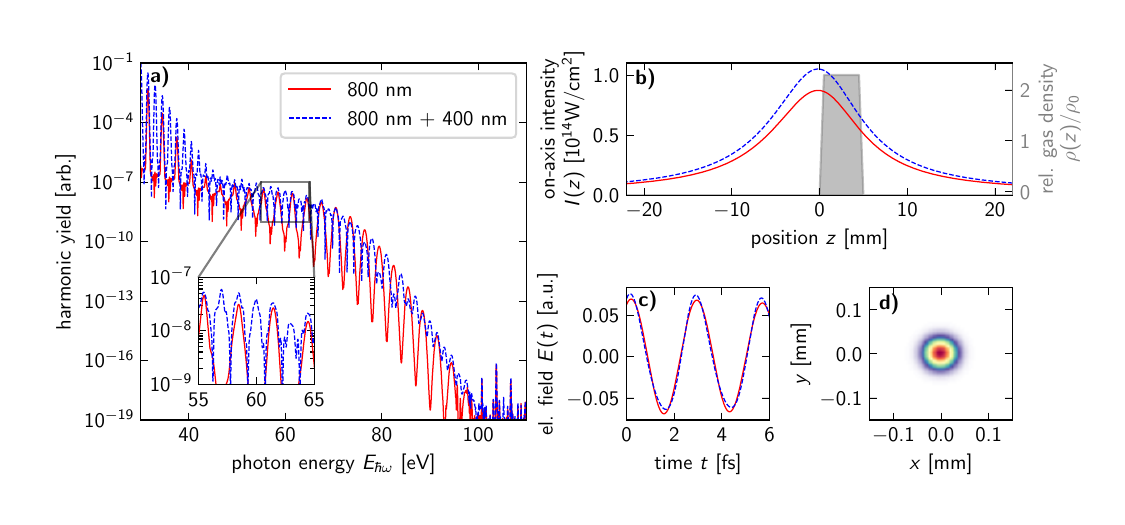}
	\caption{Simulation of two-color high-harmonic generation with a long ($\sim 25\,\mathrm{fs}$) driving pulse at $800\,\mathrm{nm}$ from a $4\,\mathrm{mm}$ long gas cell backed with $100\,\mathrm{mbar}$ of Helium with a small admixture of its second harmonic. Panel (a) shows the harmonic spectra at the end of the simulation grid (red solid line for $800\,\mathrm{nm}$ only, blue dashed for $800\,\mathrm{nm} + 400\,\mathrm{nm}$). The inset illustrates the emergence of even harmonics in the case of the two-color driving field, while for the single-color field only the expected odd-order harmonics are emitted. Panel (b) shows the on-axis intensity profile of the driving field (red solid and blue dashed lines). The gas density profile relative to the maximum density is also shown in gray.  Panel (c) shows a zoom-in onto the driving pulses at the gas cell center, the spatial profile of the harmonic emission at the end of the simulation grid for the two-color case is shown in panel (d).}
	\label{img:twocolor}
\end{figure}
Two-color driving fields for high-harmonic generation have attracted considerable attention both due to the fundamental interest in the additional control over the single-atom response as well as due to the utility of the emitted radiation in ultrafast spectroscopy experiments. Here we demonstrate that the code can be used to simulate two-color high-harmonic generation (in fact, an arbitrary number of driving wavelengths may be specified) by simulating harmonic emission from a long ($4\,\mathrm{mm}$), low-pressure ($100\,\mathrm{mbar}$) gas medium driven by a long ($\sim 25\,\mathrm{fs}$) pulse at moderate intensities ($\sim ~10^{14}\mathrm{W/cm^2}$) focused to a $40\,\mathrm{\mu m}$ radius spot. The second harmonic's field strength is set to $10\%$ of that of the fundamental. For the purpose of this demonstration we assume the gas density to increase linearly over half a millimeter, then stay constant over the length of the gas cell and decrease linearly again at its exit. More realistic density profiles can easily be implemented, if so desired. Figure \ref{img:twocolor} summarizes the result of this calculation. The emergence of even-order harmonics in the two-color case in addition to the odd-order harmonics typically observed for a single-color driving field (e.g. \cite{vuraweis2013femtosecond}) is the expected result, as validated.

\subsection{Soft x-ray high harmonic generation}
\label{ex:4}
High-harmonic sources emitting continua spanning into the soft x-rays have seen considerable interest in the recent years due to the unprecedented and detailed insight they yield into e.g. chemical dynamics (e.g. \cite{attar2017femtosecond, Bhattacherjee_2018, ridente2023femtosecond} and references therein). These sources are typically driven by longer-wavelength pulses ($1300\,\mathrm{nm}-1800\,\mathrm{nm}$ is common) and may require high intensities (on the order of $1-2\cdot 10^{15}\,\mathrm{W/cm^2})$ and gas pressures ($2-4\,\mathrm{bar})$ to operate. Under such conditions the Kerr-nonlinearity as well as the plasma generated via strong-field ionization of the nonlinear medium will strongly reshape the driving field in both space and time, presenting an additional challenge for numerical modelling.
\begin{figure}[h!]
    \includegraphics[width = \textwidth]{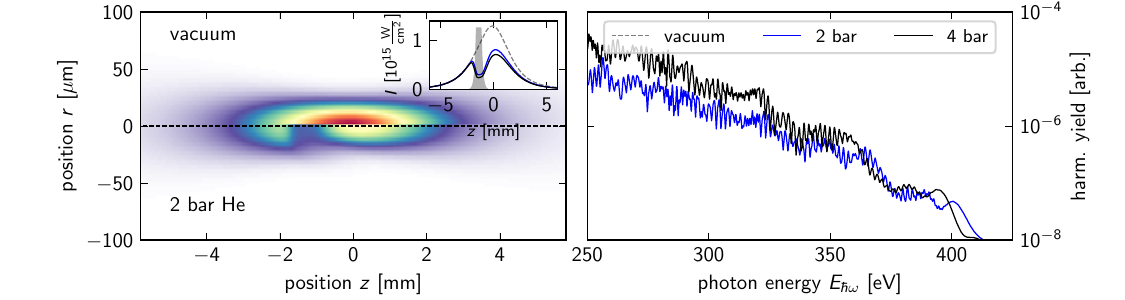}
    \caption{Soft x-ray high harmonic generation calculated with our simulation program. The left panel shows the evolution of the fundamental field in vacuum as well as in the gas medium, illustrating the strong reshaping of the driving field due to ionization, with the on-axis intensity graphed in the inset (gas density profile shaded in gray). Harmonic spectra are shown in the right panel.}
    \label{img:sxr}
\end{figure}
Here we calculate the harmonic emission of a short gass cell backed with $2\,\mathrm{bar}$ and $4\,\mathrm{bar}$ of Helium, driven by a $12\,\mathrm{fs}$ pulse (electric field envelope FWHM) centered around $1300\,\mathrm{nm}$ with a peak intensity of $1.3\cdot 10^{15}\,\mathrm{W/cm^2}$ focused to a $40\,\mathrm{\mu m}$ FWHM spot size. We adopt the gas cell pressure profile of \cite{johnson2018high}; calculated harmonic spectra as well as the evolution of the fundamental field is shown in fig. \ref{img:sxr}.

\begin{figure}[h!]
    \includegraphics[width = \textwidth]{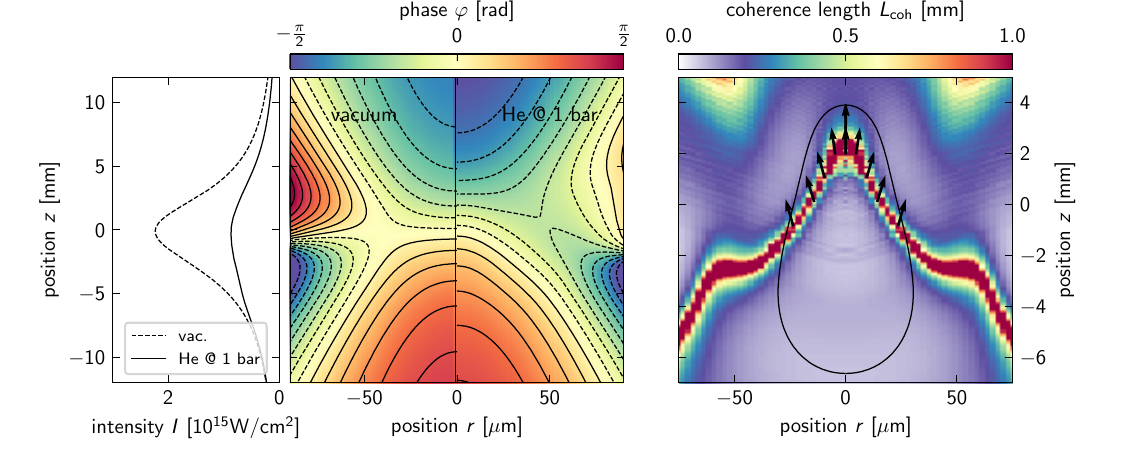}
    \caption{Intensity (left panel), phase evolution (center panel) of the driving pulses central frequency component and adiabatic map of the coherence length (right panel) calculated via eq. \ref{eq:lcoh} for generation of $300\,\mathrm{eV}$ light with a $1300\,\mathrm{nm}$ driving pulse in $1\,\mathrm{bar}$ of Helium in a semi-infinite gas cell configuration. A comparison between propagation in vacuum and the gas medium is shown in the first two panels. The strong reshaping of the driving pulse is clearly apparent by the pronounced intensity clamping. The propagation phase of the re-shaped pulse varies more quickly when compared with propagation in vacuum. Efficient generation of $300\,\mathrm{eV}$ light occurs along the thick red region in the third panel in the region where the intensity of the driving field is sufficient (encircled by the black contour line). Black arrows indicate the direction of the resulting wave vectors (eq. \ref{eq:kharm}). Optimal flux is attained when the exit of the semi-infinite gas cell is placed about $3\,\mathrm{mm}$ after the vacuum focus.}
    \label{img:lcoh}
\end{figure}
\subsection{Coherence length maps}
\label{ex:5}
In many cases a deeper understanding of the phase matching conditions can be obtained without necessarily performing a full calculation of the harmonic emission, but instead by calculating the coherence length of harmonic radiation of a given photon energy $E_\mathrm{ph}$ in a region near the fundamental beam's focus (as is done in e.g. \cite{Major2023,takahashi2003experimental}). This, however, requires the spatial evolution of the phase of the driving beam to be known. The simulation program can produce such phase output either in the spectral domain (meaning the phase evolution of a selected spectral component of the driving field), or in the time domain (meaning the phase evolution of the field at a selected time point). Here we demonstrate the spectral-domain case. With the phase $\varphi_\mathrm{d}(r, z)$ of the driving field known one can calculate the wave vector
\begin{equation}
	\vec k_\mathrm{d}(r, z) = \vec k_0 + \nabla \varphi_\mathrm{d}(r, z),
	\label{eq:kd}
\end{equation}
where $\vec k_0 = \omega_0 / c \hat e_z$. Within the Balcou model \cite{balcou1997generalized} the phase of the nonlinear polarization of the gas medium is proportional to the ponderomotive potential $U_p(r, z)$, resulting in an effective wave vector
\begin{align}
	\vec k_\mathrm{h}(r, z) = - q \tau \nabla U_p(r, z),
    \label{eq:kharm}
\end{align}
where $\tau = 5.08$ for the cut-off portion of the spectrum and $\tau = 3.16$ for the plateau region when $U_p$ is in units of electron volt \cite{balcou1997generalized}. The coherence length then is
\begin{align}
	L_\mathrm{coh}(r, z) = \frac{2\pi}{|q\vec k_\mathrm{d}(r, z) - \vec k_\mathrm{h}(r, z)|},
    \label{eq:lcoh}
\end{align}
with $q = E_\mathrm{ph} / \omega_0$ being the order of the harmonic one wishes to phase-match. The center panel in Fig. \ref{img:lcoh} shows the phase evolution of a $1300\,\mathrm{nm}$  pulse with a nominal peak intensity of $\sim 2\cdot 10^{15}\,\mathrm{W/cm^2}$ focused to a $80\,\mu\mathrm{m}$ spot at its central frequency component both when propagating in vacuum and in the $1\,\mathrm{bar}$ of Helium in a configuration where the entire focal region is filled with gas. The strong spatiotemporal reshaping of the pulse leads to pronounced clamping of the intensity (left panel) and a faster evolution of the phase as compared to the vacuum case. From eqns. \ref{eq:kd}-\ref{eq:lcoh} we can calculate the coherence length for $300\,\mathrm{eV}$ light at each $(r, z)$-point. It is shown in the right panel. The black contour line encircles the area in which the intensity and with it the ponderomotive potential is sufficient to reach this energy. Black arrows indicate the direction of the resulting wave vector (eq. \ref{eq:kharm}). This calculation can be used for designing a harmonic source e.g. based on a semi-infinite gas cell design. Placing the exit face of the gas cell around $2\,\mathrm{mm}$ after the in-vacuum focus will result in optimal outcoupling of harmonic emission at the desired energy of $300\,\mathrm{eV}$ with minimal reabsorption.

\section{Conclusion and outlook}

In summary, we present a simulation program for high-harmonic generation aimed at aiding the design of table-top XUV and soft x-ray sources driven by femtosecond laser pulses. The versatility and validity of our code is demonstrated by multiple examples spanning large parameter ranges, and show that it produces results of high quality that stand in comparison with other publications. The program is fully open-source (BSD license, source code available at \url{https://gitlab.com/caschroeder/hhg}) and therefore offers a platform for others to modify it to suit their needs or develop it further in other ways. Furthermore, we plan to add a post-processing suite for our code that will ease the analysis of the calculation results.

\section*{Acknowledgements}
Funded by the Deutsche Forschungsgemeinschaft (DFG, German Research Foundation) - 546437684. Lawrence Berkeley National Laboratory (LBNL) personnel acknowledge the support of the Gas Phase Chemical Physics Program (C.S. and S.R.L.) of the U. S. Department of Energy (DOE) Office of Science, Basic Energy Sciences (BES) Program, Chemical Sciences, Geosciences, and Biosciences Division, under contract no. DE-AC02-05CH11231 to LBNL.

\bibliography{bibliography}

\begin{thebibliography}{10}
\newcommand{\enquote}[1]{``#1''}

\bibitem{ferray1988multiple}
M.~Ferray, A.~L'Huillier, X.~F. Li, \emph{et~al.}, \enquote{Multiple-harmonic
  conversion of 1064 nm radiation in rare gases,}
  {\protect\JournalTitle{Journal of Physics B: Atomic, Molecular and Optical
  Physics}} \textbf{21}, L31 (1988).

\bibitem{li1989multiple}
X.~F. Li, A.~L'Huillier, M.~Ferray, \emph{et~al.}, \enquote{Multiple-harmonic
  generation in rare gases at high laser intensity,}
  {\protect\JournalTitle{Phys. Rev. A}} \textbf{39}, 5751--5761 (1989).

\bibitem{lewenstein1994theory}
M.~Lewenstein, P.~Balcou, M.~Y. Ivanov, \emph{et~al.}, \enquote{Theory of
  high-harmonic generation by low-frequency laser fields,}
  {\protect\JournalTitle{Phys. Rev. A}} \textbf{49}, 2117--2132 (1994).

\bibitem{balcou1997generalized}
P.~Balcou, P.~Sali`eres, A.~L'Huillier, and M.~Lewenstein, \enquote{Generalized
  phase-matching conditions for high harmonics: The role of field-gradient
  forces,} {\protect\JournalTitle{Phys. Rev. A}} \textbf{55}, 3204--3210
  (1997).

\bibitem{gaarde2008macroscopic}
M.~B. Gaarde, J.~L. Tate, and K.~J. Schafer, \enquote{Macroscopic aspects of
  attosecond pulse generation,} {\protect\JournalTitle{Journal of Physics B:
  Atomic, Molecular and Optical Physics}} \textbf{41}, 132001 (2008).

\bibitem{heyl2016scale}
C.~M. Heyl, H.~Coudert-Alteirac, M.~Miranda, \emph{et~al.},
  \enquote{Scale-invariant nonlinear optics in gases,}
  {\protect\JournalTitle{Optica}} \textbf{3}, 75--81 (2016).

\bibitem{johnson2018high}
A.~S. Johnson, D.~R. Austin, D.~A. Wood, \emph{et~al.}, \enquote{High-flux soft
  x-ray harmonic generation from ionization-shaped few-cycle laser pulses,}
  {\protect\JournalTitle{Science Advances}} \textbf{4}, eaar3761 (2018).

\bibitem{vabek_thesis}
J.~V{\'a}bek, \enquote{{Multiscale approach to the description of
  high-harmonics generation in gases},} Theses, {Universit{\'e} de Bordeaux ;
  Universit{\'e} technique tch{\`e}que. Facult{\'e} de G{\'e}nie nucl{\'e}aire
  et de Sciences de l'ing{\'e}nieur (Prague)} (2022).

\bibitem{vabek_proceedings}
J.~Vabek, S.~Skupin, and F.~Catoire, \enquote{{Multi-scale model of HHG in
  gases},} in \emph{{3rd Annual Workshop of the COST Action Attosecond
  Chemistry 2022},}  (Prague, Czech Republic, 2022).

\bibitem{hernandezgarcia2010high}
C.~Hernández-García, J.~A. Pérez-Hernández, J.~Ramos, \emph{et~al.},
  \enquote{High-order harmonic propagation in gases within the discrete dipole
  approximation,} {\protect\JournalTitle{Physical Review A}} \textbf{82}
  (2010).

\bibitem{hhgmax}
M.~Högner, \url{https://mhoegner.gitlab.io/hhgmax-homepage/} (2014).

\bibitem{Boerzsoenyi2008}
A.~B\"{o}rzs\"{o}nyi, Z.~Heiner, M.~P. Kalashnikov, \emph{et~al.},
  \enquote{Dispersion measurement of inert gases and gas mixtures at 800 nm,}
  {\protect\JournalTitle{Appl. Opt.}} \textbf{47}, 4856--4863 (2008).

\bibitem{henke1991xray}
B.~Henke, E.~Gullikson, and J.~Davis, \enquote{X-ray interactions:
  Photoabsorption, scattering, transmission, and reflection at e = 50-30,000
  ev, z = 1-92,} {\protect\JournalTitle{Atomic Data and Nuclear Data Tables}}
  \textbf{54}, 181--342 (1993).

\bibitem{lehmeier1985nonresonant}
H.~Lehmeier, W.~Leupacher, and A.~Penzkofer, \enquote{Nonresonant third order
  hyperpolarizability of rare gases and n2 determined by third harmonic
  generation,} {\protect\JournalTitle{Optics communications}} \textbf{56},
  67--72 (1985).

\bibitem{adk}
M.~V. Ammosov, N.~B. Delone, and V.~P. Krainov, \enquote{Tunnel ionization of
  complex atoms and atomic ions in electromagnetic field,} in \emph{High
  intensity laser processes,}  vol. 664 (SPIE, 1986), pp. 138--141.

\bibitem{gaarde2006spatial}
M.~B. Gaarde, M.~Murakami, and R.~Kienberger, \enquote{Spatial separation of
  large dynamical blueshift and harmonic generation,}
  {\protect\JournalTitle{Phys. Rev. A}} \textbf{74}, 053401 (2006).

\bibitem{antoine1996theory}
P.~Antoine, A.~L'Huillier, M.~Lewenstein, \emph{et~al.}, \enquote{Theory of
  high-order harmonic generation by an elliptically polarized laser field,}
  {\protect\JournalTitle{Phys. Rev. A}} \textbf{53}, 1725--1745 (1996).

\bibitem{sanderson2019practical}
C.~Sanderson and R.~Curtin, \enquote{Practical sparse matrices in c++ with
  hybrid storage and template-based expression optimisation,}
  {\protect\JournalTitle{Mathematical and Computational Applications}}
  \textbf{24}, 70 (2019).

\bibitem{sanderson2025armadillo}
C.~Sanderson and R.~Curtin, \enquote{Armadillo: an efficient framework for
  numerical linear algebra,} in \emph{2025 17th International Conference on
  Computer and Automation Engineering (ICCAE),}  (IEEE, 2025), pp. 303--307.

\bibitem{FFTW05}
M.~Frigo and S.~G. Johnson, \enquote{The design and implementation of {FFTW3},}
  {\protect\JournalTitle{Proceedings of the IEEE}} \textbf{93}, 216--231
  (2005). Special issue on ``Program Generation, Optimization, and Platform
  Adaptation''.

\bibitem{hentschel2001attosecond}
M.~Hentschel, R.~Kienberger, C.~Spielmann, \emph{et~al.}, \enquote{Attosecond
  metrology,} {\protect\JournalTitle{Nature}} \textbf{414}, 509--513 (2001).

\bibitem{chini2014generation}
M.~Chini, K.~Zhao, and Z.~Chang, \enquote{The generation, characterization and
  applications of broadband isolated attosecond pulses,}
  {\protect\JournalTitle{Nature Photonics}} \textbf{8}, 178--186 (2014).

\bibitem{vuraweis2013femtosecond}
J.~Vura-Weis, C.-M. Jiang, C.~Liu, \emph{et~al.}, \enquote{Femtosecond
  m2,3-edge spectroscopy of transition-metal oxides: Photoinduced oxidation
  state change in $\alpha$-fe2o3,} {\protect\JournalTitle{The Journal of
  Physical Chemistry Letters}} \textbf{4}, 3667--3671 (2013).

\bibitem{attar2017femtosecond}
A.~R. Attar, A.~Bhattacherjee, C.~D. Pemmaraju, \emph{et~al.},
  \enquote{Femtosecond x-ray spectroscopy of an electrocyclic ring-opening
  reaction,} {\protect\JournalTitle{Science}} \textbf{356}, 54--59 (2017).

\bibitem{Bhattacherjee_2018}
A.~Bhattacherjee and S.~R. Leone, \enquote{Ultrafast x-ray transient absorption
  spectroscopy of gas-phase photochemical reactions: A new universal probe of
  photoinduced molecular dynamics,} {\protect\JournalTitle{Accounts of Chemical
  Research}} \textbf{51}, 3203--3211 (2018).

\bibitem{ridente2023femtosecond}
E.~Ridente, D.~Hait, E.~A. Haugen, \emph{et~al.}, \enquote{Femtosecond symmetry
  breaking and coherent relaxation of methane cations via x-ray spectroscopy,}
  {\protect\JournalTitle{Science}} \textbf{380}, 713--717 (2023).

\bibitem{Major2023}
B.~Major, K.~Kov\'acs, E.~Svirplys, \emph{et~al.}, \enquote{High-order harmonic
  generation in a strongly overdriven regime,} {\protect\JournalTitle{Phys.
  Rev. A}} \textbf{107}, 023514 (2023).

\bibitem{takahashi2003experimental}
E.~Takahashi, V.~Tosa, Y.~Nabekawa, and K.~Midorikawa, \enquote{Experimental
  and theoretical analyses of a correlation between pump-pulse propagation and
  harmonic yield in a long-interaction medium,} {\protect\JournalTitle{Physical
  Review A}} \textbf{68} (2003).

\end{thebibliography}

\end{document}